\documentclass[12pt]{article}
\usepackage{amsmath,amsfonts,epsfig}

\numberwithin{equation}{section}

\begin{document}

\begin{titlepage}
\begin{flushleft}
\hfill  TIFR-TH-05-34\\
\hfill  hep-th/0512005 \\

\end{flushleft}
\vspace*{8mm}

\begin{center}

{\Large { Generalised T-Duality and Non-Geometric Backgrounds} }\\

\vspace*{12mm}

{ Atish Dabholkar$^{\rm a}$
and Chris Hull$^{\rm b}$
} \\

\vspace*{7mm}

{\em $^{\rm a}$Department of Theoretical Physics, Tata
Institute of Fundamental Research,}\\

{\em Homi Bhabha Road, Mumbai 400005, India}\\

\vspace*{4mm}

{\em $^{\rm b}$The Blackett Laboratory, Imperial College London} \\
{\em Prince Consort Road, London SW7 2AZ, United Kingdom} \\

\vspace{.4cm}
{\em $^{\rm b}$ The Institute for Mathematical
Sciences, Imperial College London,\\48 Princes Gardens, London SW7 2AZ, United Kingdom} \\

\vspace*{12mm}

\end{center}

\begin{abstract}
We undertake a systematic analysis of non-geometric backgrounds in
string theory by seeking stringy liftings of a class of gauged
supergravity theories. In addition to conventional flux
compactifications and non-geometric T-folds with T-duality
transition functions, we find a new class of non-geometric
backgrounds with non-trivial dependence on the dual coordinates that
are conjugate to the  string winding number. We argue that
T-duality acts in our class of theories, including  those cases without
isometries in which the conventional Buscher rules cannot be
applied, and that these generalised T-dualities can take T-folds
or flux compactifications on twisted tori to examples of the new
non-geometric backgrounds. We show that the new class of
non-geometric backgrounds and the generalised T-dualities arise
naturally in string field theory, and are readily formulated in
terms of a doubled geometry, related to generalised geometry. At
special points in moduli space, some of the non-geometric
constructions become equivalent to asymmetric orbifolds which are
known to provide consistent string backgrounds. We construct the
bosonic sector of the corresponding gauged supergravity theories
and show that they have a universal form in any dimension, and in
particular construct the scalar potential. We apply this to the
supersymmetric WZW model, giving the complete non-linear structure
for a class of WZW-model deformations.

\end{abstract}

\vfill

\noindent {Email: {atish@tifr.res.in}, {c.hull@imperial.ac.uk} }

\end{titlepage}






\def\be{\begin{equation}}       \def\eq{\begin{equation}}
\def\ee{\end{equation}}         \def\eqe{\end{equation}}

\def\bea{\begin{eqnarray}}      \def\eqa{\begin{eqnarray}}
\def\ena{\end{eqnarray}}        \def\eea{\end{eqnarray}}
                                \def\eqae{\end{eqnarray}}


\font\titlerm=cmr10 scaled\magstep3

\font\titlerms=cmr7 scaled\magstep3

\font\titlermss=cmr5 scaled\magstep3

\font\titlei=cmmi10 scaled\magstep3

\font\titleis=cmmi7 scaled\magstep3

\font\titleiss=cmmi5 scaled\magstep3

\font\titlesy=cmsy10 scaled\magstep3

\font\titlesys=cmsy7 scaled\magstep3

\font\titlesyss=cmsy5 scaled\magstep3

\font\titleit=cmti10 scaled\magstep3

\font\ticp=cmcsc10

\font\secfont=cmbx7 scaled\magstep3

\font\subfont=cmcsc8

\font\mysecfont=cmcsc10

\font\myfont=cmbx8

\font\mytitlefont=cmbx12

\font\eightrm=cmr8

\font\sevenrm=cmr7

\font\fourrm=cmr5

\font\cmss=cmss10



\def\a{\alpha}
\def\b{\beta}
\def\c{\gamma}
\def\d{\delta}
\def\e{\epsilon}           
\def\f{\phi}               
\def\g{\gamma}
\def\h{\eta}
\def\i{\iota}
\def\j{\psi}
\def\k{\kappa}                    
\def\l{\lambda}
\def\m{\mu}
\def\n{\nu}
\def\o{\omega}  \def\w{\omega}
\def\p{\pi}                
\def\q{\theta}  \def\th{\theta}                  
\def\r{\rho}                                     
\def\s{\sigma}                                   
\def\t{\tau}
\def\u{\upsilon}
\def\x{\xi}
\def\z{\zeta}
\def\D{\Delta}
\def\F{\Phi}
\def\Gg{\Gamma}
\def\J{\Psi}
\def\L{\Lambda}
\def\O{\Omega}  \def\W{\Omega}
\def\P{\Pi}
\def\Q{\Theta}
\def\U{\Upsilon}
\def\X{\Xi}
\def\del{\partial}              


\def\ca{{\cal A}}
\def\cb{{\cal B}}
\def\cc{{\cal C}}
\def\cd{{\cal D}}
\def\ce{{\cal E}}
\def\cf{{\cal F}}
\def\cg{{\cal G}}
\def\ch{{\cal H}}
\def\ci{{\cal I}}
\def\cj{{\cal J}}
\def\ck{{\cal K}}
\def\cl{{\cal L}}
\def\cm{{\cal M}}
\def\cn{{\cal N}}
\def\co{{\cal O}}
\def\cp{{\cal P}}
\def\cq{{\cal Q}}
\def\car{{\cal R}}
\def\cs{{\cal S}}
\def\ct{{\cal T}}
\def\cu{{\cal U}}
\def\cv{{\cal V}}
\def\cw{{\cal W}}
\def\cx{{\cal X}}
\def\cy{{\cal Y}}
\def\cz{{\cal Z}}


\def\bop#1{\setbox0=\hbox{$#1M$}\mkern1.5mu
        \vbox{\hrule height0pt depth.04\ht0
        \hbox{\vrule width.04\ht0 height.9\ht0 \kern.9\ht0
        \vrule width.04\ht0}\hrule height.04\ht0}\mkern1.5mu}
\def\Box{{\mathpalette\bop{}}}                        
\def\pa{\partial}                              
\def\de{\nabla}                                       
\def\dell{\bigtriangledown} 
\def\su{\sum}                                         
\def\pr{\prod}                                        
\def\iff{\leftrightarrow}                      
\def\conj{{\hbox{\large *}}} 
\def\lconj{{\hbox{\footnotesize *}}}          
\def\dg{\sp\dagger} 
\def\ddg{\sp\ddagger} 

\def\>{\rangle} 

\def\<{\langle} 
\def\Dsl{D \hskip-.6em \raise1pt\hbox{$ / $ } }



\def\sp#1{{}^{#1}}                             
\def\sb#1{{}_{#1}}                             
\def\oldsl#1{\rlap/#1}                 
\def\sl#1{\rlap{\hbox{$\mskip 1 mu /$}}#1}
\def\Sl#1{\rlap{\hbox{$\mskip 3 mu /$}}#1}     
\def\SL#1{\rlap{\hbox{$\mskip 4.5 mu /$}}#1}   
\def\Tilde#1{\widetilde{#1}}                   
\def\Hat#1{\widehat{#1}}                       
\def\Bar#1{\overline{#1}}                      
\def\Bra#1{\Big\langle #1\Big|}                       
\def\Ket#1{\Big| #1\Big\rangle}                       
\def\bra#1{\langle #1|}                       
\def\ket#1{| #1\rangle}
\def\VEV#1{\Big\langle #1\Big\rangle}                 
\def\abs#1{\Big| #1\Big|}                      
\def\sbra#1{\left\langle #1\right|}            
\def\sket#1{\left| #1\right\rangle}            
\def\sVEV#1{\left\langle #1\right\rangle}      
\def\sabs#1{\left| #1\right|}                  
\def\leftrightarrowfill{$\mathsurround=0pt \mathord\leftarrow \mkern-6mu
       \cleaders\hbox{$\mkern-2mu \mathord- \mkern-2mu$}\hfill
       \mkern-6mu \mathord\rightarrow$}
\def\dvec#1{\vbox{\ialign{##\crcr
       \leftrightarrowfill\crcr\noalign{\kern-1pt\nointerlineskip}
       $\hfil\displaystyle{#1}\hfil$\crcr}}}          
\def\hook#1{{\vrule height#1pt width0.4pt depth0pt}}
\def\leftrighthookfill#1{$\mathsurround=0pt \mathord\hook#1
       \hrulefill\mathord\hook#1$}
\def\underhook#1{\vtop{\ialign{##\crcr                 
       $\hfil\displaystyle{#1}\hfil$\crcr
       \noalign{\kern-1pt\nointerlineskip\vskip2pt}
       \leftrighthookfill5\crcr}}}
\def\smallunderhook#1{\vtop{\ialign{##\crcr      
       $\hfil\scriptstyle{#1}\hfil$\crcr
       \noalign{\kern-1pt\nointerlineskip\vskip2pt}
       \leftrighthookfill3\crcr}}}
\def\der#1{{\pa \over \pa {#1}}}               
\def\fder#1{{\d \over \d {#1}}} 


\def\ha{\frac12}                               
\def\sfrac#1#2{{\vphantom1\smash{\lower.5ex\hbox{\small$#1$}}\over
       \vphantom1\smash{\raise.4ex\hbox{\small$#2$}}}} 
\def\bfrac#1#2{{\vphantom1\smash{\lower.5ex\hbox{$#1$}}\over
       \vphantom1\smash{\raise.3ex\hbox{$#2$}}}}      
\def\afrac#1#2{{\vphantom1\smash{\lower.5ex\hbox{$#1$}}\over#2}}  
\def\dder#1#2{{\pa #1\over\pa #2}}        
\def\secder#1#2#3{{\pa\sp 2 #1\over\pa #2 \pa #3}}          
\def\fdder#1#2{{\d #1\over\d #2}}         
\def\on#1#2{{\buildrel{\mkern2.5mu#1\mkern-2.5mu}\over{#2}}}
\def\On#1#2{\mathop{\null#2}\limits^{\mkern2.5mu#1\mkern-2.5mu}}
\def\under#1#2{\mathop{\null#2}\limits_{#1}}          
\def\bvec#1{\on\leftarrow{#1}}                 
\def\oover#1{\on\circ{#1}}                            
\def\dt#1{\on{\hbox{\LARGE .}}{#1}}                   
\def\dtt#1{\on\bullet{#1}}                      
\def\ddt#1{\on{\hbox{\LARGE .\kern-2pt.}}#1}             
\def\tdt#1{\on{\hbox{\LARGE .\kern-2pt.\kern-2pt.}}#1}   


\newskip\humongous \humongous=0pt plus 1000pt minus 1000pt
\def\caja{\mathsurround=0pt}
\def\eqalign#1{\,\vcenter{\openup2\jot \caja
       \ialign{\strut \hfil$\displaystyle{##}$&$
       \displaystyle{{}##}$\hfil\crcr#1\crcr}}\,}
\newif\ifdtup
\def\panorama{\global\dtuptrue \openup2\jot \caja
       \everycr{\noalign{\ifdtup \global\dtupfalse
       \vskip-\lineskiplimit \vskip\normallineskiplimit
       \else \penalty\interdisplaylinepenalty \fi}}}
\def\li#1{\panorama \tabskip=\humongous                
       \halign to\displaywidth{\hfil$\displaystyle{##}$
       \tabskip=0pt&$\displaystyle{{}##}$\hfil
       \tabskip=\humongous&\llap{$##$}\tabskip=0pt
       \crcr#1\crcr}}
\def\eqalignnotwo#1{\panorama \tabskip=\humongous
       \halign to\displaywidth{\hfil$\displaystyle{##}$
       \tabskip=0pt&$\displaystyle{{}##}$
       \tabskip=0pt&$\displaystyle{{}##}$\hfil
       \tabskip=\humongous&\llap{$##$}\tabskip=0pt
       \crcr#1\crcr}}


\def\phil{@{\extracolsep{\fill}}}
\def\unphil{@{\extracolsep{\tabcolsep}}}

\def\to{\rightarrow}

\def\pa{\partial}
\def\del{\nabla}
\def\delbar{\bar{\nabla}}

\def\xx{\times}
\def\apm{\alpha^{\prime}}

\def\ab{\bar{a}}
\def\bb{\bar{b}}
\def\cb{\bar{c}}
\def\db{\bar{d}}
\def\eb{\bar{e}}
\def\fb{\bar{f}}  \def\Fb{\bar{\F}}
\def\kb{\bar{k}}  \def\Kb{\bar{K}}
\def\lb{\bar{l}}  \def\Lb{\bar{L}}
\def\mb{\bar{m}}  \def\Mb{\bar{M}}
\def\nb{\bar{n}}  \def\Tb{\bar{T}}
\def\zb{\bar{z}}  \def\Gb{\bar{G}}
\def\wb{\bar{w}}  \def\Jb{\bar{J}}
\def\pb{\bar{p}}
\def\Ib{\bar{I}}

\def\Pbb{\bar{\P}}
\def\jbb{\bar{\j}}
\def\qbb{\bar{\q}}
\def\Sbb{\bar{\S}}
\def\Pbb{\bar{\P}}
\def\dbb{\bar{\delta}}
\def\kbb{\bar{\kappa}}
\def\ebb{\bar{\epsilon}}

\def\ua{\underline{a}}
\def\ub{\underline{b}}
\def\uc{\underline{c}}
\def\ud{\underline{d}}
\def\ue{\underline{e}}

\def\uua{\underline{\a}}
\def\uub{\underline{\b}}
\def\uuc{\underline{\c}}
\def\uud{\underline{\d}}
\def\uue{\underline{\e}}
\def\uug{\underline{\g}}

\def\hL{\hat{L}}
\def\hM{\hat{M}}
\def\hj{\hat{\j}}
\def\hf{\hat{\f}}
\def\hMb{\bar{\hat{M}}}
\def\hfb{\bar{\hat{\f}}}
\def\hP{\hat{\Psi}}

\def\dif{\partial}
\def\difb{\bar{\partial}} \def\dbar{\bar{\partial}}
\def\pab{\bar{\pa}}
\def\nonu{\nonumber \\{}}
\def\half{{1 \over 2}}

\def\bfx{{\bf x}}
\def\bfy{{\bf y}}
\def\bfk{{\bf k}}
\def\bfl{{\bf l}}
\def\vp{{\vec p}}
\def\vx{{\vec x}}

\def\cn{${\bf C}/{\bf Z}_N\ $}
\def\zn{${\bf Z}_N\ $}

\def\CH{{\cal H}}
\def\CR{{\cal R}}
\def\CM{{\cal M}}
\def\CF{{\cal F}}
\def\CS{{\cal S}}
\def\CL{{\cal L}}
\def\CI{{\cal I}}
\def\CD{{\cal D}}
\def\CZ{{\cal Z}}
\def\sN{\scriptscriptstyle N}
\def\tN{{\tilde N}}
\def\ap{$\alpha '$}
\def\IZ{\relax\ifmmode\mathchoice
{\hbox{\cmss Z\kern-.4em Z}}{\hbox{\cmss Z\kern-.4em Z}}
{\lower.9pt\hbox{\cmsss Z\kern-.4em Z}} {\lower1.2pt\hbox{\cmsss
Z\kern-.4em Z}}\else{\cmss Z\kern-.4em }\fi}
\def\IC{\relax\hbox{$\inbar\kern-.3em{\rm C}$}}
\def\IR{\relax{\rm I\kern-.18em R}}
\def\bZ{{\bf Z}}
\def\bI{{\bf I}}
\def\bM{{\bf M}}
\def\bC{{\bf C}}
\def\bN{{\bf N}}
\def\bR{{\bf R}}
\def\bT{{\bf T}}
\def\bS{{\bf S}}
\def\zb{\bar z}
\def\zt{{\tilde z}}
\def\yt{{\tilde y}}
\def\xt{{\tilde x}}
\def\ft{{\tilde f}}
\def\gt{{\tilde g}}

\def\z{ z}



\def\be{\begin{equation}}
\def\ee{\end{equation}}
\def\ba{\begin{eqnarray}}
\def\ea{\end{eqnarray}}
\def\bq{\begin{quote}}
\def\eq{\end{quote}}
\def\PL{{ \it Phys. Lett.} }
\def\PRL{{\it Phys. Rev. Lett.} }
\def\NP{{\it Nucl. Phys.} }
\def\PR{{\it Phys. Rev.} }
\def\MPL{{\it Mod. Phys. Lett.} }
\def\IJMP{{\it Int. J. Mod .Phys.} }
\newcommand{\labell}[1]{\label{#1}\qquad_{#1}} 
\newcommand{\labels}[1]{\vskip-2ex$_{#1}$\label{#1}} 
\newcommand{\reef}[1]{(\ref{#1})}
\newcommand{\vs}[1]{\vspace{#1 mm}}
\def\al{\alpha}
\def\gam{\gamma}
\def\tal{\widetilde{\alpha}}
\def\sig{\sigma}
\def\part{\partial}
\def\bound{{\sigma=0,\pi}}
\def\beq{\begin{equation}}
\def\eeq{\end{equation}}
\def\beqa{\begin{eqnarray}}
\def\eeqa{\end{eqnarray}}
\def\tX{\widetilde{X}}
\def\zb{{\bar{z}}}
\def\bz{{\bar{z}}}
\def\hD{\widehat{D}}
\def\hM{\widehat{M}}
\def\Om{\Omega}
\def\F{{\cal F}}
\def\G{{\cal G}}
\def\A{{\cal A}}
\def\B{{\cal B}}
\def\tA{\tilde{\cal A}}
\def\ie{{\it i.e.,}\ }
\def\eg{{\it e.g.,}\ }
\def\iden{1\!\!1}
\def\M{{\cal K}}

\def\Z {\mathbb{Z}}
\def\R {\mathbb{R}}
\def\C {\mathbb{C}}
\def\bZ{\mathbb{Z}}
\def\bK{{\bf K}}

\section{Introduction}\label{Introduction}

Superstring theory can be formulated in \lq non-geometric'
backgrounds that cannot be understood in terms of supergravity and
conventional field theory
\cite{Hull:2004in,Dabholkar:2002sy,Flournoy:2004vn,Kachru:2002sk,Hellerman:2002ax,
Lowe:2003qy,Flournoy:2005xe,Shelton:2005cf,Gray:2005ea}, and these
offer an interesting window onto  stringy physics  beyond the scope
of supergravity. The purpose of this paper is to systematically
analyse the possible structures of non-geometric backgrounds   and
seek string-theory constructions of these.

The non-geometric backgrounds that have been studied so far are
backgrounds in which the transition functions are allowed to
include duality transformations, and a systematic study of those
with T-duality transition functions was made in
\cite{Hull:2004in}, where the name T-fold was proposed. Our
considerations will lead us to a new class of backgrounds that are
even more non-geometric than T-folds: whereas T-folds look like  conventional spacetimes locally, with ordinary spacetime patches glued together with transition functions that can include T-dualities, the new class of backgrounds do not look like  conventional spaces even locally.
 Nonetheless, we shall
present arguments that these non-geometric spaces are good string backgrounds and that
they necessarily arise in string theory. In some cases, these new
backgrounds are obtained from consistent backgrounds by the action
of T-dualities. However, these are T-dualities in directions in
which there are no isometries, so that the conventional
sigma-model duality transformations through the Buscher rules
cannot be applied. We argue that T-duality can nonetheless be
applied, with a more general set of rules.

A classic example is that of a three-torus with constant $H$-flux,
so that the NS-NS 3 form field strength is proportional to the
volume form on $T^3$. Performing one T-duality gives a $T^2$
bundle over $S^1$ \cite{Hull:1998vy} with monodromy in the group
$SL(2,\Z)$ of large diffeomorphisms on $T^2$. A further T-duality
on one of the fibre directions gives a non-geometric T-fold that
is a $T^2$ bundle over $S^1$, but with monodromy in the T-duality
group $O(2,2;\Z)$ \cite{Kachru:2002sk, Lowe:2003qy, Hull:2004in},
so that this should be thought of as a bundle of CFTs over a
circle with each fibre being  a CFT on $T^2$ and the moduli of
these CFTs varying over the $S^1$. A further T-duality on the $S^1$
base appears impossible as the T-fold geometry depends explicitly
on the circle coordinate. We propose a construction of  the missing T-dual.

This has interesting applications to mirror symmetry on Calabi-Yau
manifolds with flux. Consider a CY with $T^3$ fibration with
$H$-flux on the fibres. If the argument of
\cite{Strominger:1996it} generalises to the case with flux, then
one expects   the mirror background in the large volume limit to
be obtained by acting with T-dualities on all fibre directions. In
this case, the T-dual must be a non-geometric background of the
type we propose here.

We show that the new non-geometric backgrounds and the generalised
T-dualities arise naturally in string field theory, and indeed
generic solutions of string field theory will be non-geometric
backgrounds of this type. We also show that in some cases the new
exotic backgrounds   have a point in moduli space that minimises a
scalar potential and at which the theory can be constructed by a
special type of asymmetric orbifold, so that our new constructions
can be thought of as giving deformations of asymmetric orbifolds.

The new backgrounds arise naturally in the doubled formalism used
in  \cite{Hull:2004in}. Strings on a torus $T^n$ can be formulated
on a doubled torus $\bar T^{2n}$ with $n$ coordinates $X^i$
conjugate to the momenta and $n$ coordinates $\tilde X^i$
conjugate to the winding modes. Geometric backgrounds and T-folds
can have non-trivial dependence on the space coordinates $X^i$ in
the usual way, while our new backgrounds have non-trivial
dependence on $\tilde X$ also. For example, a $T^n$ bundle over
$T^m$ gives a geometric background, a $\bar {T}^{2n}$  bundle over
$T^m$ gives a T-fold while bundles of the doubled torus  $\bar
T^{2n}$  over the doubled base $\bar T^{2m}$ give examples of our
new class of non-geometric backgrounds. Indeed, the T-dual of the
$T^3$ with $H$-flux is a bundle of this kind with $n=2,m=1$.

Our approach here is to investigate the lifting of a certain class
of $D$-dimensional gauged supergravities to $10$ dimensions. In
some cases they arise from supergravity compactifications, but for
others there is no such geometric origin. Some   arise from
intrinsically stringy constructions such as asymmetric orbifolds,
reductions with duality twists or T-fold reductions.  In general,
the lifting is non-geometric in the sense that there is no
supergravity compactification that gives rise to them. For
example, compactification on a manifold with an isometry group $K$
gives a theory in which the vector fields from the reduction of
the metric are gauge fields for a Yang-Mills group $K$, but the
gauge fields arising from the reduction of antisymmetric tensor
gauge fields typically have abelian  self-interactions, so that
the gauge group for the reduction of a theory with gravity plus anti-symmetric tensors  is a product  or semi-direct product of $K$ with an abelian
group. There are gauged supergravities in which the
$D$-dimensional gauge fields that are associated with
antisymmetric tensor gauge fields become non-abelian in $D$
dimensions, but  such non-abelian interactions cannot arise from a
conventional compactification. Roughly speaking, the curving of
the internal geometry leading to an isometry group gives a
non-abelian structure to the momentum modes, and what would be
needed is the dual version of this that gives a non-abelian
structure to string winding modes or brane wrapping modes, and
this is clearly beyond the scope of supergravity. The
non-geometric background we find gives  the required \lq curving'
of the dimensions conjugate to the string winding modes

An important question is whether all effective supergravity
theories have a stringy lifting.  We find classical stringy
liftings for a wide class of \lq unliftable' supergravities and we
have found no obstructions to finding such lifitings generally.
However, in this paper we consider only classical string theory,
and at the quantum level modular invariance is expected to
restrict the possible non-geometric backgrounds, just as it does
for asymmetric orbifolds. However, those that arise as T-duals of
consistent backgrounds should be quantum consistent.

We focus here on  the sector associated with the metric, B-field
and dilaton, and on the action of T-duality, as we are interested
here in explicit string theory constructions, but this can be
generalised to include U-duality and RR or heterotic sectors. The
dimensional reduction of this sector on a $d$-torus and truncation
to a particular finite subset of fields gives a theory with an
$O(d,d)$ symmetry, $2d $ vector fields and $d^2+1$ scalars. The
gauged theories we consider are ones in which a $2d$-dimensional
subgroup of $O(d,d)$ is promoted to a local symmetry with the spin
one fields becoming  the gauge fields. We argue that the general
gauged supergravities must be of the same form as the ones found
by Kaloper and Myers \cite{Odd} for Scherk-Schwarz reductions with flux,
generalised to a wider class of gauge groups than considered
there. In particular, we find the form of the scalar potential for
the $d^2+1$ scalars.

The effective $D$-dimensional supergravity theories that we
consider give important information about the structure of the
10-dimensional liftings, when they exist, and provide a way of
parameterising the possible backgrounds that might emerge. In
general, these arise from a (possibly non-geometric)
compactification followed by a truncation. There is a natural
action of $O(d,d)$ on the effective supergravity, which takes it
to an apparently different but physically equivalent
$D$-dimensional theory.  However, these transformations can have
radical effects on the 10-dimensional liftings, for example taking
one from a geometric compactification in which the non-abelian
gauge fields arise from isometries to a non-geometric one   in
which the non-abelian vector fields come from the $B$-fields
coupling to winding modes. This gives an indication  as to how
T-duality must act in the uplifted theory. The scalar potential
also gives important information about the structure of the
theories. Of particular interest are critical points  of the
scalar potential with zero energy, as they correspond to
reductions to $D$-dimensional Minkowski space.  An important class
of these arise from asymmetric orbifold constructions, so that the
non-geometric deformations away from the orbifold point in moduli
space that we propose correspond to  moving away from the minimum
of the potential, and  there will in general no longer be a
solution involving $D$-dimensional Minkowski space. In these
cases, the solution of the $D$-dimensional effective theory which has maximal symmetry is typically a solution of
domain-wall type with $D-1$ dimensional Poincar\' e symmetry.

The plan of this paper is as follows. In $\S{\ref{Action}}$ we
review the low energy effective action for toroidal
compactifications of string theory and consider the most general
gaugings. Scherk-Schwarz reductions and duality-twisted reductions
 are special cases of these. Based on the
known formula for the effective potential that arises in this
case, we propose an effective potential for the most general
gauging based on the duality covariance of the formula and discuss
some properties of the general gauge algebra in
$\S{\ref{algebra}}$. We review and discuss Scherk-Schwarz reductions and
duality-twisted reductions  in sections $\S{\ref{general}}$ and
$\S{\ref{Dualitytwist}}$ respectively. In $\S{\ref{WZW}}$, we apply our formalism to WZW
models and  argue that the potential
proposed in $\S{\ref{Action}}$  correctly captures the non-linear
structure for a class of WZW-model deformations. In particular,
the potential gives the correct symmetry breaking
pattern after including an infinite number of higher order terms
in the effective potential for the Higgs field that breaks the
gauge symmetry. In $\S{\ref{New}}$ we show
that some of the models obtained by lifting  gaugings are
related   by duality to known Scherk-Schwarz and duality-twisted
reductions, and investigate in detail the
 special cases in which these reduce to asymmetric orbifold constructions which can be dualised explicitly.
   In $\S{\ref{nongeometric}}$ we examine the new
non-geometric constructions implied by the generalised T-duality.
In $\S{\ref{sft}}$ we argue that the generalised T-duality can be
implemented within string field theory and
that general string field theory solutions will  include non-geometric backgrounds of the type we discuss here.
  We conclude with a discussion in
$\S{\ref{discussion}}$.

While this paper was in preparation, the paper
\cite{Shelton:2005cf} appeared in which the relation between
low-energy effective actions in four  dimensions and non-geometric
backgrounds were also explored. They also identified   classes of
gaugings that might lead to new non-geometric backgrounds, but no
constructions  of such backgrounds  were provided.

\section{The Low-Energy Effective Action and Gauged Supergravity} \label{Action}

The low-energy effective field theory for $n$-dimensional string
theory includes the terms
\begin{equation}
S = \int d^{n}x \sqrt{-{\cal G}} e^{- \Phi} \Bigl\{ {\cal R}
 +  (\nabla \Phi)^2 - \frac{1}{12} {\cal H}_{\mu \nu \lambda}
{\cal H}^{\mu \nu \lambda}\Bigr\} \label{sact1}
\end{equation}
for a scalar field $\Phi$, metric ${\cal G}_{\mu\nu}$ and 2-form
gauge field ${\cal B} $ with field strength  ${\cal H} = d{\cal
B}$. Our conventions  are as in \cite{Odd}, and our notation
largely follows that of \cite{Odd}. This action can arise as part
of the low-energy effective action of the bosonic string with
$n=26$ or of the \lq common sector'  of the heterotic or type II
superstrings in $n=10$ dimensions. The 10-dimensional supergravity
theories describing the low-energy limit of the  heterotic or type
II string theories can be decomposed into $D=10,N=1$ multiplets.
In all cases, there is a common sector consisting of the $N=1$
supergravity multiplet whose bosonic fields consist of the metric,
the 2-form gauge field ${\cal
B}$ and the dilaton, with action
(\ref{sact1}). For the heterotic or type I strings, there are in
addition Yang-Mills multiplets, while for the type II strings one
adds a gravitino multiplet whose highest spin state is a
gravitino, and there are two possibilities depending on whether
this has the same or the opposite chirality to the gravitino in
the supergravity multiplet. We will  discuss in this paper the
reduction of the sector of the low-energy theory represented by
the action (\ref{sact1}), corresponding to the $N=1$ supergravity
sector of the theory, obtaining a gauged supergravity with $16$
supersymmetries that is a truncation of the gauged supergravity
arising from the full theory. The reduction of the full heterotic
theory is discussed in \cite{Odd}, while that of the type II
theory will be discussed in \cite{reida}. In particular, the
scalar potential we obtain will be  restricted to the scalars
coming from the reduction of the common sector. However, in the
full theory there are other scalars coming from the RR fields in
the type II string and from the Yang-Mills gauge fields in the
heterotic string, and the full potential   depends on these also.
The gauge symmetry of the full theory will contain the gauge
algebra that we discuss in section 3, but will be larger in
general. The generalisation to M-theory   will
be discussed in \cite{reida}.

The standard Kaluza-Klein dimensional reduction of
(\ref{sact1}) on a $d$-torus gives a theory in $D=n-d$ dimensions
with an $O(d,d)$ invariance  and action \cite{actor, Odd} \ba S
&=& \int d^{D} x \sqrt{-{g}} e^{-\phi} \left\{{R}
 + ({\nabla} \phi)^2 - \frac{1}{12} {H}^2_{\mu\nu\lambda}\right.
\nonumber\\
&&\left.\qquad\qquad\qquad + \frac{1}{8}
 L_{ab} \nabla_\mu \M^{bc}L_{cd}\nabla^\mu \M^{da}
- \frac{1}{4} {F}^a_{\mu\nu}
 L_{ab}\M^{bc}L_{cd} { F}^{d\,\mu\nu}\right\}
\label{actodd} \ea The $d$ Kaluza-Klein vector fields
$V^M{}_{\mu}$ ($M,N=1,...,d$) from the reduction of the metric and
the $d$ vector fields $B_{\mu M}$ from the reduction of ${\cal B}$
have been combined into the $2d$ vector fields $A^a_{\mu}$
($a,b=1,...,2d$) with abelian  field strengths $F^a=dA^a$
\begin{equation}
A^a_{\mu}~=~\begin{pmatrix} V^M{}_{\mu}\\
                          B_{\mu M}
                          \end{pmatrix}
\label{vectm}\end{equation} These vector fields transform  as the
fundamental representation of $O(d,d)$, $A^a\rightarrow
{A'}^a=M^a{}_b A^b$. The $O(d,d)$ invariant metric is \be
L^{ab}=~\begin{pmatrix}
~0~&{\bf 1} ~ \\ {\bf 1}&~0~
  \end{pmatrix}
\label{ldef} \ee
where ${\bf 1}$ is the  $~ d \times d~$ unit matrix, and satisfies
$M\,L\,M^T=L$, where the superscript $T$ indicates
matrix transposition. The scalar fields take values in the coset
$O(d,d)/O(d) \times O(d)$ and are represented by a symmetric
$2d\times 2d$ matrix $\M^{ab}$ with trace $L_{ab}\M^{ab}=0$ and
which transforms under $O(d,d)$ as $\M^{ab}\rightarrow
\M'^{ab}=M^a{}_cM^b{}_d\M^{cd}$.
 The reduced
three-form field strength is \begin{equation}
 {H}_{\mu\nu\lambda} = \partial_{\mu}
{B}_{\nu\lambda} - \frac{1}{2} {A}^a_{\mu}\, L_{ab}\,
{F}^b_{\nu\lambda} + cyclic ~permutations
\label{axfsin}\end{equation} and includes abelian Chern-Simons
terms \cite{MS}. The theory has $U(1)^{2d}$ gauge symmetry and a
manifest global $O(d,d)$ symmetry.

We will be interested in gauged supergravities that arise as
deformations of this theory. As we shall review in
$\S\ref{general}$, a class of these arise from dimensional
reduction on twisted tori with flux. In the class of gauged supergravities that  we shall
consider here, a $2d$-dimensional subgroup $G$ of $O(d,d)$ is
promoted to a local symmetry, with the gauge fields $A^a$
appearing in the minimal couplings. For this to be possible, it is
essential that the fundamental representation of $O(d,d)$ becomes
the adjoint of $G$ under the embedding of $G$ in $O(d,d)$, so that
the gauge fields $A^a$ transform in the adjoint of the gauge
group. If the generators of $O(d,d)$ are denoted by
$t_{ab}=-t_{ba}$, then the subgroup $G$ generators $T_a$ are
specified by an embedding  tensor $\Theta _{a}{}^{bc}$ so that \be
T_a = \frac{1}{2} \Theta _{a}{}^{bc} t_{bc} \ee and satisfy
\be [T_a, T_b] = i f_{ab}{}^c T_c \label{algebra0} \ee
where
$f_{ab}{}^c $ are the structure constants of $G$.

The deformation of (\ref{actodd})  with gauge symmetry $G$ is
\begin{eqnarray}
S &=& \int d^{D} x \sqrt{-{g}} e^{-\phi} \Bigl\{{R}
 + ({\nabla} \phi)^2 + \frac{1}{8}
L_{ab} {\cal D}_\mu \M^{bc} L_{cd} {\cal D}^\mu \M^{da}
 \nonumber \\
&&~~~~~~~~~~~~~~~~~~~~~ - \frac{1}{4} F^a_{\mu\nu}
 L_{ab} \M^{bc} L_{cd} F^{d\mu\nu}
- \frac{1}{12} {H}^2_{\mu\nu\lambda} - g^2 W(\M) \Bigr\}
\label{actoddna1}
\end{eqnarray}
where $W(\M)$ is a gauge-invariant scalar potential, the covariant derivative
of the scalar fields is
\begin{equation}
{\cal D}_\mu \M^{ab} = \partial_\mu \M^{ab} -g f_{cd}{}^a A^c_{\mu
} \M^{db} - g f_{cd}{}^b A^c_{\mu} \M^{ad}\ . \label{covdermf1}
\end{equation}
the field strengths for the gauge fields $A$ taking values in the
Lie algebra of $G$ are
\begin{equation}
F = d A + i g A \wedge A = \frac12 F^a_{\mu\nu} T_a dx^\mu \wedge
dx^\nu \label{gfla1}
\end{equation}
and the   three-form field strength is \be H = dB - \frac12
\omega_{CS} \ee where the Chern-Simons term for $G$ is \be
\omega_{CS} =tr( A \wedge F - g\frac{i}3 A \wedge A \wedge A)
\label{csfin1} \ee
Here $g$ is the gauge coupling constant, which will be set to $g=1$ in later sections.
In the supersymmetric case, this is the bosonic sector of a gauged supergravity
action in which the fermionic terms are modified by fermion mass terms proportional
to $g$.

All the terms in the action (\ref{actoddna1}) are completely
determined by gauge invariance and requiring the ungauged limit to
give (\ref{actodd}), apart from the scalar potential, which in
principle could be any $G$-invariant function of the scalars in
the bosonic string theory. However, for the superstring, the
theory resulting  from the reduction of the supergravity action
whose bosonic part contains (\ref{actoddna1}) is a gauged
supergravity in $D$ dimensions and supersymmetry leads to further
restrictions. Supersymmetry in general   rules out some of the
possible gauge groups, and determines the form of the scalar
potential. The minimal couplings for the class of the theories
considered here only depend on the embedding tensor through the
structure constants $f_{ab}{}^c$, and on general grounds the
supersymmetric scalar potential must be $O(d,d)$ covariant and
must be constructed from the structure constants $f_{ab}{}^c$, the
scalar matrix $ \M^{ab} $ and the $O(d,d)$ metric $ L^{ab}$.
Moreover, it must be quadratic in the structure constants and of
order $g^2$ and so the general form must be (using the notation
   $f_{abc} =f_{ab}{}^dL_{cd}$, which is automatically antisymmetric $f_{abc} =f_{[abc]} $)
\begin{equation}
W_{gen}(\M)= a \M^{ad} \M^{be} \M^{cf} f_{abc} f_{def} +b \M^{ad} \M^{be}L^{cf}
f_{abc}f_{def}
+c
 \M^{ad} L^{be}L^{cf} f_{abc}f_{def}
 +d
  \label{potcgen}
\end{equation}
for some constants $a,b,c,d$, and it must take this form with the
same constants $a,b,c,d$ for all possible gaugings.

In the generalised Scherk-Schwarz reductions with flux to be discussed in $\S{\ref{general}}$, the scalar
potential is given by $W(\M)=  W_{SS}(\M)$ \cite{Odd,reid} where
\begin{equation}
W_{SS}(\M)= \frac{1}{12} \M^{ad} \M^{be} \M^{cf} f_{abc} f_{def} -
\frac{1}{4} \M^{ad} L^{be}L^{cf} f_{abc}f_{def}\ . \label{potcov1}
\end{equation}
This must agree with the general potential, and is sufficiently general to fix
the  coefficients $a,b,c,d$, and so the potential must be $W(\M)=
 W_{SS}(\M)$  given by (\ref{potcov1}) for all supersymmetric gaugings.
Our discussion of explicit string compactifications and their low-energy limits
will provide non-trivial checks of this conclusion.
Note that in general there could be quantum corrections to this potential.

The ungauged action is manifestly invariant under $O(d,d)$
transformations, while the couplings of the gauged action
(\ref{actoddna1}) break this to the subgroup preserving the
structure constants $f_{ab}{}^c $. However, it becomes invariant
if the structure constants are also taken to transform \be
f_{ab}{}^c  \to M_a{}^d M_b {}^e M^c{}_ff_{de}{}^f
\label{ftrans}
\ee so that the
embedding also changes.

\section{The Gauge Algebra and Duality}\label{algebra}

As we shall review in the next section, generalised Scherk-Schwarz
reduction is specified by two constant tensors,  the structure
constants $\gamma^M_{NP}$ of a Lie group $K$  specifying the \lq
twisting' of the basis one-forms $\eta^M$ in (\ref{formalg2}), and
the constraints $\beta_{MNP}$ specifying a 3-form flux
\begin{equation}
 \frac12 \beta_{MNP} \eta^M \wedge \eta^N \wedge \eta^P
\end{equation}
for the 3-form field strength ${\cal H}$.

As we have seen, the $2d $ gauge fields $A^a$ consist of $d$ gauge
fields $V^M$ arising from the reduction of the metric and $d$
gauge fields $B_M$ from the reduction of the B-field. The gauge
generators $T_a$ decompose into the generators $Z_M$ corresponding
to $V^M$  and $X^M$ corresponding to $B_M$. The internal space has
isometry group $K$, and in the absence of flux the gauge group is
a semi-direct product of the group containing $K$ generated by $Z_M$ and an
abelian group generated by the $X^M$. With flux, this is deformed
to \cite{Odd,reid} \ba &&[X^M, X^N] = 0 \nonumber \\
&&[X^M, Z_N] = 2 i \gamma^M_{NP} X^P \nonumber \\ &&[Z_M, Z_N] =
2i \gamma^P_{MN} Z_P-3i \beta_{MNP} X^P
 \label{algebra1} \ea
and the low-energy effective action is given by (\ref{actoddna1})
and (\ref{potcov1}) based on this algebra. The Jacobi identities
require that $\gamma^M_{NP}$ are the structure constants of a Lie
algebra $K$ and that \be 3 \beta_{L[MN} \gamma^L_{PQ]} =
0\label{mbetagammaa} \ee

The most general gauge algebra will have a decomposition
\ba &&[X^M, X^N] =  2i \tilde \g ^{MN}{}_PX^P + 3i \tilde \b ^{MNP}Z_P \nonumber
\\ &&[X^M, Z_N] = 2 i \hat  \gamma^M_{NP} X^P+ 2i  \bar \g ^{MP}_NZ_P \nonumber \\
&&[Z_M, Z_N] =  2i \gamma^P_{MN} Z_P-3i \beta_{MNP} X^P
\label{algebra11} \ea
and is specified by a set of tensors $\g,\b,\tilde \g...$, subject
to the constraints arising from the Jacobi identities.

These then   constitute the general set of parameters specifying
the low-energy field theory generalising the twist $\g$ and flux
$\b$ of the usual reduction. We will refer to $\tilde \g $ as the
dual twist and $\tilde \b$ as the dual flux. Our aim here is to
find string constructions in which some of these new
parameters $\tilde \g,\hat  \gamma, \bar \g ,\tilde \b$ are non-trivial.
   In some cases, these arise from
   non-geometric backgrounds, but as we shall see in $\S{\ref{WZW}}$, they can also arise from geometric compactifications that are not of Scherk-Schwarz type.

As we saw in the last section, there is a natural action of
$O(d,d)$ under which the structure constants $f_{ab}{}^c$
transform covariantly. These transformations   mix the twist $\g$
and flux $\b$ with each other and with the new
parameters $\tilde \g,\hat  \gamma, \bar \g ,\tilde \b$  in general.
This gives a natural conjecture for the action of the T-duality
group $O(d,d;\Z)$ in the full string theory \cite{reid}, namely
that the structure constants transform as (\ref{ftrans}). This
reproduces the known cases in which T-duality acts through the
Buscher rules \cite{buscher} and  interchange twist and flux (e.g.
acting on a 3-torus with constant H-flux $\b$ and $\g=0$ with a
T-duality on one circle gives a torus bundle over a circle with
twist $\g$ but no flux, so that $\b=0$ \cite{Hull:1998vy}). Moreover, this predicts how to
extend this to non-geometric backgrounds.

\section{  Scherk-Schwarz Reductions with Flux} \label{general}

In this section we review the results of \cite{Scherk:1979zr,Odd,reid} for
the general Scherk-Schwarz reduction with flux of the theory with
action (\ref{sact1}) to $D$ dimensions. The internal manifold is
taken to be a space with a basis of one-forms $\eta^M$ satisfying
\be d \eta^M = - \gamma^M_{NP} \eta^N \wedge
\eta^P \label{formalg2}
\ee
with constant coefficients $\gamma^M_{NP}$, so that the
integrability condition for (\ref{formalg2}) is that the
$\gamma^M_{NP}$ satisfy the Jacobi identity
\be \gamma^M_{R[N}\,
\gamma^R_{PQ]} = 0 \label{congamma2}
\ee
and so are the structure constants of the Lie algebra for  some group $K$.
Note that $K$ can be non-compact or non-semi-simple. The
Scherk-Schwarz reduction ansatz  for tensor fields is that they
have components with respect to the $\eta$-frame that are
independent of the internal coordinates $y$, so that the internal
metric is
\begin{equation}
ds^2= {\cal G}_{MN} \eta^M\eta^N \label{g1ansatz}
\end{equation}
where $ {\cal G}_{MN}$ is independent of the internal coordinates
$y$. The $ {\cal G}_{MN}$ are moduli for the internal metric and
in the full ansatz can depend on the remaining coordinates
$x^\mu$, giving rise to scalar fields ${\cal G}_{MN}(x)$.
Similarly, the internal B-field could be taken to be ${\cal B}
=\frac12 {\cal B}_{MN} \eta^M \wedge \eta^N$. This can be
generalised to include a flux $ \frac12 \beta_{MNP} \eta^M \wedge
\eta^N \wedge \eta^P$ for some  constants $\beta_{MNP}$. This will
be a  closed 3-form  if
 \be  \beta_{L[MN} \gamma^L_{PQ]} =
0\label{mbetagamma} \ee
Then locally there is a 2-form $\omega $
so that \begin{equation} d\omega =  \frac12 \beta_{MNP} \eta^M
\wedge \eta^N \wedge \eta^P  \label{gansatz}
\end{equation}
and the full ansatz for the 2-form can be written as
\begin{equation}
 {\cal B} =\frac12 {\cal B}_{MN}  \eta^M \wedge \eta^N +\omega
 \label{bansatz}
\end{equation}
Thus the construction is specified by constants $\gamma^M_{NP}$,
$\beta_{MNP}$ satisfying (\ref{congamma2}),(\ref{mbetagamma}) and
has internal moduli ${\cal G}_{MN},  {\cal B}_{MN}$ giving $d^2$
scalar fields in the compactified theory,  in addition to the dilaton, just as for the torus
case (the constants $\gamma^M_{NP}$, $\beta_{MNP}$ do not give
further moduli). This is sufficient to ensure that the reduced
field equations have no dependence on the internal coordinates,
but reduction of the action requires  that the volume element can
also be reduced to give an action that is independent of the
internal coordinates, and this requires the further condition
\be \gamma^N_{NM} = 0\ .\label{congamma3} \ee

The construction is very similar to a toroidal compactification,
but with the coordinate frame $dy^M$ replaced by the $\eta^M$
frame, which is  \lq twisted' by the constants $\gamma^M_{NP}$ to
give what is sometimes referred to as a \lq twisted torus' associated with the  $d$-dimensional
group $K$. For reduction on an  internal space with metric and B-field given by  (\ref{g1ansatz}) and
(\ref{bansatz}),  Scherk and Schwarz  gave an ansatz that allows
the truncation of the $n$-dimensional field theory to a
$(D=n-d)$-dimensional one that is independent of the internal
coordinates $y$. However, for this to be a compactification, one
needs an internal space that is compact with the local structure
(\ref{formalg2}). If the isometry group $K$ is compact, we can
take the internal manifold to be the group manifold $K$ with
isometry group $K_L\times K_R$ with the $\eta^M$ the
left-invariant Maurer-Cartan one-forms. In general, the internal
space must be the quotient $K/\Gg$ of the group manifold by a
discrete subgroup of $\Gg$ of $K_L$, the left action of $K$ on
itself \cite{reid}. If $K$ is non-compact, it is necessary to
choose $\Gg$ so that $K/\Gg$ is compact in order for the
Scherk-Schwarz reduction and truncation of a field theory to be
extendable to a compactification of string theory. A necessary
condition  \cite{reid} for the existence of such a $\Gg $ is
   (\ref{congamma3}). We will assume (\ref{congamma3}) here.

The ansatz for the reduction is the most general one that is
invariant under the left action $K_L$. For supergravity, the
reduction can be viewed as reduction on the group manifold
followed by truncation to the $K_L$ singlet sector,  and the
isometry of the internal space $K_R$ gives rise to a gauge group
$K$ \cite{reid}. For the full string theory, we wish to consider
the full theory without truncation, so we  require   a compact
internal space. For  $K$ non-compact, we consider reduction on
compact spaces $K/\Gamma$ where $\Gamma$  is a  discrete subgroup
of $K_L$. The left-invariant one-forms $\eta $ on $K$ then give
well-defined one-forms on $K/\Gamma$.

Let $x^\mu$ be $D$-dimensional space-time coordinates and  $y^M$
be coordinates on the $d$-dimensional internal space. It is useful
to define
 \begin{equation}
\nu^M=\eta^M(y) + V^M{}_\mu(x) dx^\mu
\end{equation}
where $V^M{}_\mu(x)$ are the $d$ Kaluza-Klein vector fields which
are gauge fields for the isometry group $K$, with field strength
$dV^M +\gamma^M_{NP}V^N\wedge V^P$. The full ansatz for the reduction of
the metric is \cite{Scherk:1979zr}
\be ds^2 = g_{\mu\nu}(x) dx^\mu
dx^\nu + {\cal G}_{MN} (x)\nu^M\nu^N
 \label{nonabkkmet} \ee
where
 the metric ${\cal G}_{MN}(x)$
gives $d(d+1)/2$ scalar fields.
The
ansatz for the 2-form gauge field is
\begin{equation}
{\cal B} = \frac12 {\cal B}_{\mu\nu} dx^\mu \wedge
dx^\nu \wedge  +  {\cal B}_{\mu M} dx^\mu
\wedge  \nu^M
+\frac12 {\cal B}_{MN}  \nu^M \wedge \nu^N +\omega
\end{equation}
with
3-form field strength $ {\cal H}=d{\cal B}$.

The reduced theory has $2d$ gauge fields
$A^a_\mu=(V^M{}_\mu,B_{\mu M})$. The $V^M{}_\mu$ are gauge fields
corresponding to  the isometry group $K_R$ of $K$, and in the
standard case in which there is no flux ($\beta =0$) the
corresponding generators of gauge transformations $Z_M$ satisfy
the $K$ commutation relations $[Z_M, Z_N] =   2i \gamma^P_{MN}
Z_P$. There are a further $d$ gauge symmetries for which $B_{\mu
M}$ are the gauge fields, with generators $X^M$. The full  reduced
theory has a $2d$-dimensional gauge group $G$ generated by the
$Z_M,X^M$. The algebra was calculated in \cite{Odd}
and found to be \ba &&[X^M, X^N] = 0 \nonumber \\ &&[X^M, Z_N] = 2
i \gamma^M_{NP} X^P \nonumber \\ &&[Z_M, Z_N] = -3i \beta_{MNP}
X^P + 2i \gamma^P_{MN} Z_P \label{algebra3} \ea so that the flux
modifies the algebra. The Jacobi identities are satisfied as a
result of (\ref{congamma2}) and (\ref{mbetagamma}). If $K$ is
semi-simple and the flux is given by the structure constants, then the term involving the flux $ \beta_{MNP}$ can
be removed by redefinitions \cite{Cvetic:2003jy}, but in general this is not
possible  \cite{reid}.

Introducing the generators $T_a = (Z_M, X^M)$ of $G$, the algebra
can be written as
\be [T_a, T_b] = i f_{ab}{}^c T_c \label{algebra4} \ee
where
\begin{equation}
f^M{}_{NP} =f_{NP}{}^M= 2 \gamma^M_{NP} ~~~~~~~~~~ f_{MNP} = -3
\beta_{MNP} \label{strc1}
\end{equation}
The gauge group $G$ is a subgroup of the $O(d,d)$ that was a
symmetry of the toroidally reduced theory.
The reduced action is (\ref{actoddna1}).

\section{Reductions with Duality Twists}\label{Dualitytwist}

There is a second type of dimensional reduction which is also
often called a Scherk-Schwarz reduction. If a theory in $D+1$
dimensions has a  global symmetry $K$, then reducing on a
further circle to $D$ dimensions, one can allow a twist around
that circle by an element of $K$. We   refer to these as
reductions with duality twists \cite{Dabholkar:2002sy}, and will show that while a large
class of these are Scherk-Schwarz reductions of the type
considered in the previous section, this is not true in general,
and the twisted reductions which are not of Scherk-Schwarz type
will play an important role. We will be interested here in the
case in which a string theory is reduced first on a $d-1$ torus to
give a theory in $D+1$ dimensions with global symmetry
$O(d-1,d-1;\Z)$ and reducing on a further circle with an
$O(d-1,d-1;\Z)$ twist. The low-energy field theory in $D+1$
dimensions in fact has an $O(d-1,d-1)$ continuous symmetry and in
the field theory one could twist with an element of $O(d-1,d-1)$,
while in the full string theory there is an action of $O(d-1,d-1)$
on the theory, but only the discrete subgroup $O(d-1,d-1;\Z)$  is a symmetry and
the  twist must be in this discrete subgroup. This is the typical
situation in the examples arising in string theory; there is an
action of a continuous group $K$ of which a discrete subgroup
$K(\Z)$ is a symmetry, and the ansatz uses the action of $K$
but the twist is required to be in $K(\Z)$ \cite{Hull:1998vy}.

Suppose an element $g$ of a continuous   group $K$ in $D+1$
dimensions acts on a generic field $\psi$ as $\psi \to g[\psi]$.
For our string theory case, $K=O(d-1,d-1)$. Consider now the
twisted dimensional reduction of the theory to $D$ dimensions on a
circle of radius $R$ with a periodic coordinate $y\sim y+ 2 \pi
R$. In the twisted reduction, the fields are not independent of
the internal coordinate but are chosen to have a specific
dependence on the circle coordinate $y$ through the ansatz
\be { \psi(x^{\mu}, y) = g(y) \, [ \psi (x^{\mu})] }
\label{twistansatz}
\ee
for some $y$-dependent group element $g(y)\in K$. An important
restriction on $g(y)$ is that the   reduced theory in $D$
dimensions should be independent of $y$. This is achieved by
choosing
\be { g(y)= \exp \left( \frac{My}{2\pi R} \right) }
\label{Mtwist}
 \ee
for some Lie-algebra element $M$. The map $g(y)$ is not periodic
around the circle in general, but has a {\it monodromy}
\be { {\cal M} (g)= \exp M.} \ee
In the
string theory, the monodromy has to be an element of
${\cal G}(\Z)=O(d-1,d-1)(\Z)$ \cite{Hull:1998vy}.

The reduction for string theory with an $O(d-1,d-1;\Z)$ twist  is
straightforward and gives a $D$ dimensional theory with effective action
(\ref{actoddna1}) and a non-semi-simple gauge group $G$. The coordinates $y^M$
split into the coordinates $y^m$ on the $(d-1)$-torus with
$m=1,..,d-1$ and the coordinate $y$ on the final circle,
$y^M=(y,y^m)$, so that the gauge field one-forms are $V^y,V^m$ and
$B_y, B_m$ with corresponding gauge group generators
$Z_y,Z_m,X^y,X^m$. It will also be useful to introduce
  indices $\alpha,\beta=1,...,2(d-1)$ for the fundamental representation of
  $O(d-1,d-1)$ so that the gauge generators
  are
  $T_a=(Z_y,X^y,T_\alpha)$ where
$T_\alpha=(Z_m,X^m)$. Then the gauge algebra for this reduction is
\be [Z_y, T_\alpha] = M_\alpha {}^\beta T_\beta
\label{algebra2a}
\ee
with all other commutators vanishing, and from this the structure
constants $f_{ab}{}^c$ can be read off. The scalar potential is
\cite{Scherk:1979zr}
\be { V(\Phi) = e^{a \phi}{\rm Tr} [ M^2 + M^t K MK^{-1}],}
\label{twistpot}
\ee
where $e^\phi$ is the modulus corresponding to the radius of the
final circle and $a=6/(D-1)(D-2)$.

In the basis in which $T_\alpha=(Z_m,X^m)$, the mass matrix $M_\alpha
{}^\beta$ has the decomposition
\be M_\alpha {}^\beta=~  \begin{pmatrix}
~w_m{}^n~&u_{mn}~ \\
v^{mn}
  &~-(w^t)^m{}_n~ \end{pmatrix}\label{mdef}
\ee
where $u_{mn}=-u_{nm}$ and $v^{mn}=-v^{nm}$ while $w_m{}^n$ is an unconstrained matrix. The algebra
(\ref{algebra0}) can then be written as
 \ba &&[Z_y, Z_m] =
w_m{}^nZ_n +u_{mn}X^n \nonumber \\ && [Z_y, X^m] = -w_n{}^mX_m
+v^{mn}Z_n \label{algebra2}
 \\ &&[Z_n, Z_m] =0 \qquad  [X^m, X^n] = 0 \nonumber
\ea
This is of the same form as (\ref{algebra11}) only if $v^{mn}=0$.
Thus the twisted reduction considered here is more general than
the Scherk-Schwarz reduction of the type considered in
$\S\ref{general}$ whenever $v^{mn}$ is nonzero. Conversely, not
all Scherk-Schwarz compactifications can be viewed as
duality-twisted compactifications. (For example, Scherk-Schwarz
compactification on a non-abelian semi-simple group manifold $H$  without flux will
result in the $Z_m$ generating the Lie algebra of $H$
which is not of the form  (\ref{algebra2}) in general.)

Comparing with the general algebra (\ref{algebra11}) in
$\S{\ref{algebra}}$ we see that in this way we obtain a gauging
where $\tilde \gamma$ and $\tilde \beta$ are zero and all other
structure constants are nonzero. The Scherk-Schwarz (twisted torus) reductions
have non-zero $\g=\hat \g,\b$, but  the reduction with duality twist has a more
general algebra in which $\bar \g$ is also non-zero. We show in
the next section how   $\tilde \gamma$ and $\tilde \beta$ can
arise in the context of non-geometric shifts and asymmetric
orbifolds.

The mass matrix (\ref{mdef}) is in the Lie algebra of
$O(d-1,d-1)$. The subalgebra parameterised by $ w_m{}^n$ is
$GL(d-1,\R)$ which acts geometrically on the torus $T^{d-1}$ while
$u_{mn}$ parameterises the transformations that shift the
$B$-field on the torus by a constant 2-form. Thus the subalgebra
in which $v^{mn}=0$ is the one that can be realised geometrically
as diffeomorphisms of $T^{d-1}$ and shifts of the $B$-field and
such twists can be realised as standard reductions on a manifold
which can be viewed either as a $T^{d-1}$ bundle over a circle
with suitable $B$-field ansatz, or as a reduction on a discrete
identification of the group manifold for the group $G$ with Lie
algebra (\ref{algebra2}). When $v^{mn}\ne 0$, the T-duality twist
is non-geometrical and it is this non-geometrical case that is not
of the Scherk-Schwarz form of $\S\ref{general}$, giving an
internal space which is a T-fold \cite{Hull:2004in} rather than a
geometrical background.

\section{WZW Models}\label{WZW}

We now discuss an example where the full nonabelian structure of
the gauge group plays a role. Following the general discussion in
$\S{\ref{Action}}$, compactification of the action  (\ref{sact1}) on   $\bT^3$
gives a low energy theory  (\ref {actodd}) with an $O(3,3)$ rigid symmetry and
six vector fields  in the vector representation $(\bf{3}, \bf{3})$
of $O(3, 3)$. We can choose to gauge an $SO(3) \times SO(3)$
subgroup since the vector representation $(\bf{3}, \bf{3})$ of
$O(3, 3)$ decomposes into the adjoint $(\bf{3}, \bf{1}) + (\bf{3},
\bf{1})$ of $SO(3) \times SO(3)\subset O(3, 3)$, to give a gauged theory with action  (\ref {actoddna1}). There is a
natural candidate for the string-theory lifting of this gauging,
which is one where the internal CFT is the $SO(3)$ WZW model (or a
non-conformal deformation of this that preserves $SO(3) \times
SO(3)$ global symmetry) as this would give the same gauge symmetry
$SO(3) \times SO(3)$. We will now analyse this case in more
detail, showing that the symmetry breaking patterns emerging from
the scalar potential are exactly those of the WZW model,
supporting our claim that the WZW model is the correct lifting of
this gauging. It is remarkable that our potential (\ref{potcov1})
then provides a complete non-linear form for the potential for the
WZW model; we will explore this in more detail elsewhere \cite{inprep}.

From the embedding of  $SO(3) \times SO(3)\subset O(3, 3)$, the
generators $T_M$ and $\tilde T_M$ of the two $SO(3)$ factors
($M=1,2,3$) are related to the generators $Z_M,X^M$ by
\be
 T=Z+X, \qquad \tilde T=Z-X
\ee
(with indices raised or lowered with $\delta _{MN}$). Then the
gauge algebra takes the form
\ba &&[X^M, X^N] =  3i  \e ^{MNP}Z_P \nonumber \\ &&[X^M,
Z_N] = 2 i   \e ^M_{NP} X^P \nonumber \\ &&[Z_M,
Z_N] =  2i \e^P_{MN} Z_P  \label{algebraWZW}
\ea
Comparing with \ref{algebra11}, we see that this gives $\tilde \b
^{MNP}= \e ^{MNP}$, $  \gamma^M_{NP}=\hat  \gamma^M_{NP}=\e
^M_{NP}$. Note that gauge algebras with nontrivial $\tilde \b,
\hat  \gamma^M_{NP}$ are in this case arising from a geometric
reduction on the $SO(3)$ group manifold (or $S^3$) with flux.

The scalar fields in the $O(3,3)/O(3)\times O(3)$ coset space can
be parameterised by a matrix $\varphi^{\a\a'}$ where $\alpha = 1,
2, 3$ takes values in the adjoint representation of one  $SO(3)$
and $\alpha' = 1, 2, 3$ takes values in the adjoint representation
of the other $SO(3)'$. It is easy to see that  the potential
(\ref{potcov1}) has a leading term which is cubic in
$\varphi^{\a\a'}$ and takes the form
\begin{equation}\label{cubicpot}
W(\M) =  -4+   \epsilon_{\a\b\c} \epsilon_{\a'\b'\c'}
\varphi^{\a\a'}\varphi^{\b\b'}\varphi^{\c\c'} +O  (\varphi ^4)= -4+ \det (\varphi)+O  (\varphi ^4)
\end{equation}
The full potential is non-polynomial in  $\varphi^{\a\a'}$. In
this leading-order approximation, $\varphi$ can be thought of as
Higgs field that transforms in the bi-adjoint representation of
the gauge symmetry and thus to start with there are nine scalar
fields (plus a dilaton). The full symmetry is restored at $\varphi
=0$. The potential allows for a single flat direction and the
moduli fields can acquire a nonzero expectation value $
\varphi^{\a\a'}= \delta^{\a3} \delta^{\a'3} v$. When the vacuum
expectation value $v$ is nonzero, then the symmetry is broken to
$SO(2) \times SO(2)$. By expanding around this vev, it is easy to
see that out of the nine scalar fields, one remains massless for
all values of $v$ corresponding to the flat direction. We will
show in \cite{inprep} that this is an exact flat direction of
the full non-polynomial potential. There are four
would-be goldstone bosons that are eaten by the four gauge fields
which then acquire a mass of order $v$. Of the remaining four, two become
massive and two become tachyonic.

This symmetry breaking pattern is exactly what one would obtain by
compactifying the bosonic string on a three-dimensional $SO(3)$
group manifold with   three-form flux given by the invariant
3-form so  that it corresponds to a WZW model. Compactifying  on
the $SO(3)$ $WZW$ model one would obtain $SO(3)_L \times SO(3)_R$
as the spacetime gauge symmetry. If $J^a$ and $\tilde J^a$ are the
currents of $SO(3)_L$ and $ SO(3)_R$ respectively then there is an
exactly marginal operator in the worldsheet conformal field theory
$ \int  a J^3 \tilde J^3$ where $a$ is a real coefficient. As a
result, there is precisely one flat direction (up to gauge equivalence). The moduli space of
this compactification is one-dimensional,  labelled by this real
number $a$ which we can identify with $v$ above. Moreover, for
nonzero $a$, the addition of this perturbation breaks the symmetry
to $SO(2)_L \times SO(2)_R$ generated by $J_3$ and $\tilde J_3$.

We are interested here in the large $k$ limit of the WZW model
that is nearly classical. However,  the symmetry-breaking pattern does
not depend on  $k$, so a quick way to verify this pattern is
to look at the WZW model with $k=1$ described by a single boson
$X$ near the self-dual radius. If we   compactify the bosonic
string on a circle at the self-dual radius,  there is one flat
direction for the spacetime potential  corresponding to changing
the radius. Giving a nonzero vev to the modulus field corresponds
to moving slightly away from the self-dual radius which breaks the
symmetry to $SO(2)_L \times SO(2)_R$ generated by the worldsheet
currents $\partial X$ and $\tilde \partial X $. It is easy to see
the spectrum of states is exactly the same as above.

It is thus nontrivial that  for the natural generalisation of the
Scherk-Schwarz potential that we have proposed, the quadratic term
is absent and the potential correctly captures the symmetry
breaking pattern expected from string theory. Note also that the
vacuum energy $g^2W$ at the minimum of the potential does not vanish but instead has a small
negative value that goes as $ -4g^2$. This acts as a
tadpole for the dilaton and the equations of motion of the dilaton
would not be satisfied unless we adjust the total central charge
to be 26. The gauge coupling constant $g^2$ corresponds to $1/k$
and hence this correction to the central charge goes as $-1/k$.
Recall that the central charge of the $SO(3)$ model is
$\frac{3k}{k +2}$ which for large $k$ goes as $ 3 - \frac{6}{k} +
\ldots$. The correction $- \frac{6}{k}$ is the leading $\alpha'$
correction. It is interesting that our generalized potential is
able to capture this subleading correction even though we are
working in the framework of supergravity.

In superstring theory, we   compactify the theory on a super WZW
model at level $k$ which factorizes in terms of three
Majorana-Weyl fermions and a bosonic WZW model at level $k-2$.
Hence, our analysis above applies without much change.
Note that now the corrections to the central charge is exactly
$-\frac{6}{k}$ and there are no further $\alpha'$ corrections.

To obtain a supersymmetric, tachyon-free compactification of the
Type-II string one can consider, for example, ${\bf AdS_3} \times
{\bf S^3} \times {\bf K3} $ that arises as a near-horizon limit of
the NS5-F1-P black hole. The $\bf S^3$ factor is then described by
a super WZW model. The bulk supergravity  has a dual description
in terms of a boundary sigma model with  $(4, 4)$ superconformal
symmetry. In the boundary theory, the $SU(2)  \times SU(2)$
symmetry of the bulk appears as the R-symmetry. By turning on
various operators in the boundary CFT that break the $(4, 4)$
superconformal symmetry  to $(2, 2)$ superconformal symmetry one
can reduce the R-symmetry from $SU(2) \times SU(2)$ to $U(1)
\times U(1)$. These will correspond to interesting renormalisation
group flows in the boundary CFT. In the bulk supergravity, the
description of  this  symmetry breaking pattern far away from the
symmetric point would require the full nonlinear structure of the
effective potential above that encapsulates the higher order
interactions.

\section{T-Folds and Asymmetric Orbifolds}\label{New}

We have seen that a particular class of  gaugings (those with
gauge algebra (\ref{strc1})) arise from Scherk-Schwarz reductions
with flux, while another class (with gauge algebra
(\ref{algebra2a})) arises from reductions with duality twists, but
these constitute only a restricted class of possible gaugings. For
the remainder of this paper, we wish to investigate the question
of which other gaugings can have a string theory origin, and we
will consider in particular stringy generalizations of the
Scherk-Schwarz reduction and orbifold constructions, which are
closely related. Many of our considerations will apply equally to
the bosonic, heterotic and type II strings. Our initial goal  will
be to explore the classical structure of such a generalization
using symmetry considerations. We will see that our proposed
generalization encompasses the duality-twisted reductions along
with Scherk-Schwarz reductions  and thus includes non-geometric
possibilities that are classically consistent. We postpone
questions of quantum consistency such as modular invariance for
future investigation.

The starting point is then a choice of gauge group $G$ which is a
$2d$ dimensional subgroup of $O(d,d)$ with structure constants
$f_{ab}{}^c$ such that the fundamental representation of $O(d,d)$
becomes the adjoint of $G$ under the embedding of $G$ in $O(d,d)$.
Then the truncation to the common sector of the bosonic action in $D$
dimensions is given by
(\ref{actoddna1}) with the definitions (\ref{covdermf1}),
(\ref{gfla1}), (\ref{csfin1}).

By construction, the potential of gauged supergravity
(\ref{potcov1}) proposed in  $\S\ref{Action}$ gives  the
Scherk-Schwarz potential when the structure constants take the
form (\ref{algebra1}). It is also easy to see that when the
structure constants take the form (\ref{algebra2a}), the
potential reduces to the potential (\ref{twistpot}) of the
duality-twisted compactifications.  Note
that the duality twists with nonzero $v^{mn}$ in (\ref{algebra2})
are more general than the Scherk Schwarz type because the
T-duality group $O(d, d, \bZ)$ contains additional information
about winding modes that goes beyond supergravity. For example, as
shown in \cite{Dabholkar:2002sy}, when the duality twist is a
general element of $O(d-1, d-1, \bZ)$ that is not an element of
the geometric of $SL(d-1, \bZ)$ of the torus, the theory at the
minimum of the potential is described by an asymmetric orbifold.
Such a stringy twist clearly goes beyond the usual geometry
underlying Scherk-Schwarz reductions.

\subsection{T-dual Shifts}

We now show that the generalization (\ref{algebra11}) for the
gauge algebra proposed in $\S{\ref{algebra}}$ is realized
naturally in the context of general orbifolds. We have seen that
one can generate a class of non-geometric twists from  geometric
ones by acting with $O(d-1,d-1,\Z)$ T-duality transformations. To
go further and generate more general gauge algebras it is natural
to consider acting with $O(d,d,\Z)$ T-duality transformations. In
particular, it is natural to consider T-dualising the circle with
coordinate $y$ that carries the monodromy. However, there is an
obvious and immediate obstruction to this: translation in the
circle direction is not an isometry as the ansatz
(\ref{twistansatz}) introduces explicit $y$-dependence into the
background so that the moduli of the internal torus are
$y$-dependent. T-duality as normally formulated requires that
translation in the  direction in which one dualises to be an
isometry. However, as we shall review below, for elliptic twists
there is a point in the moduli space at which $y$-translations
become an isometry \cite{Dabholkar:2002sy} and in this case one
can T-dualise. Moreover, we showed in \cite{Dabholkar:2002sy} that
at such special points in moduli space, the  twisted reduction
becomes equivalent to an orbifold by a
$\Z_n$ symmetry  of the torus CFT together with an order $n$     shift
along a spectator circle \cite{Dabholkar:2002sy}. As we shall see, the T-dual is again an orbifold, but
now by a symmetry involving a shift in the T-dual
circle. In both cases we can consider the effective
compactified supergravity theory and the special points in moduli
space correspond to minima of the scalar potential. In the case
considered in \cite{Dabholkar:2002sy} with shifts in the
original circle, the potential gives information about deformations
away from the special point in moduli space, and it is natural to conjecture that
this applies for the new orbifold with a T-dual shift as well.
This would be remarkable, as it is unclear how to formulate string
theory in such circumstances -- the background would be
non-geometric and of a kind more general than the T-folds proposed
in \cite{Hull:2004in}. We now proceed to discuss this construction
in more detail.

The relation between duality-twisted reductions and orbifolds was
explored in \cite{Dabholkar:2002sy}. The duality twist is said to
be of the `elliptic' type if the matrix $M$ in
(\ref{Mtwist})
is a compact generator of
$O(d-1,d-1)$ and is thus conjugate to an element of the compact
subgroup $O(d-1) \times O(d-1)$. One can show that for an elliptic
twist the monodromy ${\cal M}$ is always of finite order, say $n$,
so that ${\cal M}^n=1$. This is basically because a discrete
rotation is of finite order. It was shown in
\cite{Dabholkar:2002sy} that for duality twists which are of the
elliptic type, the theory at the minimum of the potential is given
by an orbifold. The monodromy matrix ${\cal M}$ generates a $\Z_n$
subgroup of the discrete duality group. Generically, the duality
group maps one point of the moduli space to another point.
However, the minimum of the potential occurs at a value of the
moduli which is left fixed by this $\Z _n$ group. At this point,
then, the $\Z _n$ that leaves it invariant is a proper symmetry of
the theory at that point. One can therefore construct an orbifold
utilizing this $\Z _n$ symmetry. The orbifold action in this case
works as follows. Consider an $n$-fold cover of the circle and let
the coordinate on this larger circle be $Y$ (so that effectively
$Y=ny$) with periodicity $2\pi R_0 $ with $R_0 = n R$. Now,
consider an orbifold action that combines the order $n$ twist
generated by ${\cal M}$ with an order $n$ shift along this larger
circle, i.e. a shift by \be Y \to Y +2\pi R_0/n . \ee  In the the
theory over the larger circle,  all fields are periodic going
around the coordinate $Y$. The orbifolding results in  a
nontrivial bundle going around the original circle coordinate $y$
with monodromy $\cal M$.

One obvious way to generalise this structure is then to T-dualise
along the circle direction. T-duality takes the circle with radius
$ R_0 $ and coordinate $Y$ to a dual circle with radius $\tilde
R_0 =1/R_0$ and coordinate $\tilde Y \sim \tilde Y + 2\pi \tilde
R_0$. The $\bZ_n$ orbifold in the dual description is  then
generated by an element $\alpha$ corresponding to the same duality
twist as before but now accompanied by an order $n$ shift in the
T-dual circle coordinate $\tilde{Y}$ \be  \tilde Y \to \tilde Y +
2\pi \tilde R_0/n.\ee Given the T-duality symmetry of the
underlying string theory, one can always carry out the T-duality
along the $Y$ circle in the orbifold conformal field theory.
Moreover, modular invariance of   the original orbifold ensure the
modular invariance of  the T-dual orbifold.

To understand the meaning of the  shift along $\tilde Y$ more
concretely, recall that after compactifying along the circle in
the $Y $ direction, the momenta and winding along this circle take
values in a lattice $\Gamma^{1,1}$. A basis  for  the string
Hilbert space includes the states  $\ket{m, w}$ that carry $m$
units of quantized momentum and wrap $w$ times along $Y$. Now, an
order $n$ shift along $Y$ acts on these states as
\be \ket{m, w}\rightarrow \exp{(2\pi i m /n)} \ket{m, w}. \label{shift}
\ee
Under T-duality, the quantum numbers $(m, w)$ are related to the
corresponding $(\tilde{m}, \tilde{w})$ quantum numbers in the dual
theory by $m= \tilde{w}$ and  $w =\tilde{m}$. Hence, an order $n$
shift along the T-dual $\tilde{Y}$ coordinate acts on these states
as
\be \ket{\tilde{m}, \tilde{w}}\rightarrow \exp{(2\pi i\tilde{w}
/n)} \ket{\tilde{m}, \tilde{w}}. \label{dualshift} \ee
This defines the $\tilde Y$ shift in the CFT.

A simple nontrivial supersymmetric example of such an orbifold is
the following. Consider Type-II string theory on $\bf T^4 \times
S^1$. One can consider a $\Z _2$ orbifold action that acts as an
inversion on all coordinates of the $\bf T^4$ accompanied by an
order two shift along the circle. Without the shift, one would
obtain Type-II string on (the  $\bf T^4/\Z_2  \times S^1$ orbifold
limit of) $\bf K3 \times S^1$. However, with the shift, the
twisted states are massive and one will get a theory with the same
number of supersymmetries as the $\bf K3 \times S^1$
compactification but different massless spectrum. Note that the
order two shift could be either be of type (\ref{shift}) or
(\ref{dualshift}) with $n=2$, depending on whether one is shifting
along the $Y$ circle or along the the dual $\tilde Y$ circle. In
both cases, the resulting orbifold is modular invariant and
quantum consistent.

Note, however that the orbifold CFT exists only at the minimum of
the potential and not away from it.  At the minimum of the
potential, the moduli fields are at the fixed point of the
discrete symmetry. As a result, one can show that with the ansatz
(\ref{twistansatz}), the moduli fields are independent of the
coordinate $Y$ or $y$. One way to understand this fact is to note
that the Scherk-Schwarz potential arises essentially from the
gradient energy of the fields with the ansatz (\ref{twistansatz}).
At the minimum, there is no gradient energy, which means that the
fields are independent of $y$. In this case we have then
translation invariance along $y$, which allows us to apply the
usual T-duality rules.

When we move away from the minimum of the potential, the moduli of
the theory are no longer at the fixed point of the discrete
symmetry generated by $\cal M$. In this case, there is a
nontrivial dependence on the coordinate $y$ for the fields given
by the ansatz (\ref{twistansatz}). Now, without translation
invariance along $y$, we can no longer use the naive T-duality
rules. Thus, even though the theories at the minimum are related
to each other by T-duality one would expect that the theories away
from the minimum are not directly related by a simple T-duality
but in a certain sense exchange the dependence on momentum modes
with the dependence on winding modes.

\subsection{Asymmetric Twists and Shifts}

The next level of  generalization would be to allow for a shift in
both $Y,\tilde Y$ directions and in this way obtain a gauge
algebra in (\ref{algebra11}) where all $\gamma$, $\beta$, $\tilde
\gamma$, and $\tilde \beta$ are non-vanishing. The gauge algebra
in this case would be
\be [Z_y, T_\alpha] = M_\alpha {}^\beta T_\beta \ee and \be [X^y,
T_\alpha] = \tilde M_\alpha {}^\beta T_\beta. \ee
 with
 \be
 [M, \tilde M]=0
 \ee
If $M$ and $\tilde M$ are distinct but mutually commuting and
generate subgroups of order $n$ and $m$ respectively then the
theory at the minimum would be described by an $\bZ_n \times
\bZ_m$ orbifold. The orbifold action is generated by an element
$\alpha$ corresponding to an  $M$ twist accompanied by an order
$n$ shift along the $y$ circle and $\beta$ corresponding to an
$\tilde{M}$ shift accompanied by a order $m$ shift along
$\tilde{Y}$. Note  that even though that the theory at the minimum
of the potential will be described by a regular asymmetric
orbifold, in the theory away from the minimum, the fields would
have unusual non-geometric dependence on both winding and momentum
modes.

Asymmetric orbifolds are severely constrained by modular
invariance and generically an asymmetric twist and a shift would
not lead to a modular invariant theory \cite{Narain:1986qm}. Our
analysis here is entirely classical and the quantum consistency of
the theory is not guaranteed at this level, so that quantum
consistency will lead to extra constraints. However, any
background that is T-dual to a  consistent background will be
consistent. Moreover, there do exist several nontrivial examples
of orbifolds with asymmetric twists and shifts that are  quantum
consistent and which cannot be obtained  as a T-dual of a
symmetric orbifold. See, for example, \cite{Dabholkar:1998kv}.
Thus, the general algebra (\ref{algebra11}) that we have used as
the basis of our formalism of gauged supergravity can certainly be
realized within string theory at the minimum of the potential.

\section{ Generalised T-duality and New Non-Geometric Constructions}\label{nongeometric}


For reductions with duality twists, conventional T-duality cannot
be  applied to the circle on which the theory is twisted. In this
section, we generalise the standard T-duality to obtain T-duals of
configurations that have no conventional T-dual, obtaining new
classes of non-geometric backgrounds. We will argue that these must be
good string backgrounds. In this section, we will use the picture
\cite{Hull:2004in} in which taking a T-dual corresponds to change
of choice of polarisation in a doubled formalism. We will then
show in $\S{\ref{sft}}$ how this emerges naturally from string field theory.

Consider a  reduction with duality twist, with a reduction on
$T^{d-1}$ followed by reduction on a circle with an
$O(d-1,d-1;\Z)$ T-duality twist. The dependence on the coordinate
$y$ of the final circle is through the  $O(d-1,d-1)$ group element
\be { g(y)= \exp \left( \frac{My}{2\pi R} \right) }
\label{gy}
 \ee
with monodromy
\be { {\cal M} (g)= \exp M.} \ee
in $O(d-1,d-1;\Z)$. Taking the T-dual on one of the circles in
$T^{d-1}$  is straightforward as there is no dependence on the
coordinates of $T^{d-1}$. For example, for geometric twists  with
monodromy in $SL(d-1,\Z)$, the reduction is equivalent to
reduction on a $d$-dimensional compact space which is a $T^{d-1}$
bundle over $S^1$ \cite{Hull:1998vy}. The translations in the
fibre directions are isometries and so T-duality along a fibre
direction can be done according to the standard Buscher rules. In
some cases this gives another geometric reduction, while in others
it gives a T-fold \cite{Hull:2004in}. However, T-duality along the
$y$-direction is problematic, as the background depends explicitly
on $y$. In the geometric reduction, the torus bundle has no
isometry in the $y$ direction so that the Buscher rules cannot be
applied.

However, there is   evidence that it should be possible to take a
T-dual in the $y$ direction. Firstly, the circle has both momentum
and winding degrees of freedom, although the momentum $p_y$ is not
constant along the circle and depends on $y$. Nonetheless one
might expect an equivalent formulation with momentum and winding
interchanged, now with the \lq winding charge' depending on the
circle coordinate. As we shall see, a natural way this might be
achieved is by a simple  rewriting with $y$ exchanged with its
dual coordinate $\tilde y$.

Secondly, for elliptic twists, there are points in moduli space in
which the reduction becomes independent of $y$, and becomes
equivalent to an orbifold with a shift in the $y$-direction. This
can then be dualised to give an orbifold with a dual shift in the
$\tilde y$ direction. Before dualising, we know that there is no
obstruction to moving away from the special  orbifold point in
moduli space, but there is an energy cost, as the moduli become
scalar fields of the dimensionally reduced theory, and moving way
from the orbifold point involves moving away from the minimum of
the scalar potential. However, the T-dual orbifold is the same
conformal field theory, but written in different variables. This
means that it must be possible to move away from the orbifold
point in moduli space in the T-dual theory also. Moreover, this
deformation is readily understood in the dimensionally reduced
effective field theory. We have reviewed how the T-duality acts on
the effective field theory in $D$ dimensions, giving a new
effective potential. However, the T-duality can be viewed as a
change of variables to a T-dual set of fields, so that the new
potential can be thought of as being the old one written in the
new variables. Moving away from the minimum of the potential is
clearly possible in the reduced theory, again at the cost of
increase in potential energy.

Our potential (\ref{twistpot})  allows us to go away from the
minimum in the low-energy effective theory and gives a way of
thinking about the off-shell duality. The key question  is whether
one can lift this to the full theory, allowing a deformation to a
new kind of non-geometric background. One would expect that the
T-dual of an orbifold with $Y$-shift (recall that $Y=ny$ is the
coordinate on the $n$-fold cover of the circle), which is an
orbifold with a shift  along $\tilde Y$,  is a minimum of a
potential which is obtained by gauging the algebra that is T-dual
to the algebra that was gauged to obtain (\ref{twistpot}). Under
T-duality in the $y$ direction, the element $Z_y$ is conjugated to
the element $X_y$, $X_y = T Z_y T^{-1}$. Thus, in the T-dual
description when we are shifting along $\tilde Y$ direction, the
gauge algebra for this reduction is expected to be
\be [X^y, T_\alpha] = \tilde M_\alpha {}^\beta T_\beta \ee
with all other commutators vanishing.  This is obtained simply by
acting on the algebra (\ref{algebra2a}) by the T-duality
transformation along the $y$ direction. In this case, starting
with a situation with non-vanishing $\gamma$ and $\beta$ but
vanishing $\tilde \gamma$ and $\tilde \beta$ in our general gauge
algebra (\ref{algebra11}), one would obtain a gauge algebra with
non-vanishing $\tilde \gamma$ and $\tilde \beta$ but vanishing
$\gamma$ and $\beta$.

If moving away from the special point in moduli space for the
orbifold with $y$-shift gives a twisted reduction with a twist in
the $y$ direction, the natural guess  is that doing the same for
the orbifold with dual $\tilde y$-shift should give a twisted
reduction with a twist in the $\tilde y$ direction. That is, in
reducing on the circle, dependence on the dual coordinate $\tilde
y$ is introduced  through the twist
\be { g(\tilde y)= \exp \left( \frac{M\tilde y}{2\pi \tilde R} \right) }
\label{gyt}
\ee
so that the reduction is of the same form as before, but with $y$
and $\tilde y$  interchanged.  This appears rather trivial from
the conformal field theory viewpoint, but the change to a
background with dependence on the  dual coordinate takes us away
from the realm in which there is a local spacetime description.

It is useful to consider these constructions from the point of
view of conformal field theory. A state in the conformal field
theory on the circle is characterised by momentum $n/R$ and
winding number $w$ for integers $n,w$ and   is represented by the
tensor product of a  state
\be \ket{n, w}\ee
with a state in the oscillator Fock space. One can instead perform
a Fourier transform and represent a state in a position basis with
two periodic coordinates $y, \tilde y$ of periods $2\pi R$ and
$2\pi \tilde R= 2\pi/R$ respectively \be \ket{y,\tilde y}= \sum
_{n,w}e^{iny/R}e^{iw\tilde y/\tilde R}\ket{n, w} \ee T-duality
interchanges $n$ with $w$ or $y$ with $\tilde y$. In the conformal
field theory, $n$ and $w$ or  $y$ and $\tilde y$ are on an equal
footing, and can be viewed as living on the doubled space, the
2-torus with coordinates $y$ and $\tilde y$. To make contact with
the conventional spacetime picture, one must choose a
polarisation; one picks one of the two coordinates to be the
spacetime coordinate appearing in the sigma-model action, and the
other to be the dual coordinate \cite{Hull:2004in}. T-duality acts
to change the polarisation from the one in which $y$ is a
spacetime coordinate to the one in which $\tilde  y $ is. The
statement that T-duality is a symmetry implies that the physics is
independent of the choice of polarisation and that both give the
same physical results.

The conformal field theory on the torus $T^{d-1}$ is completely
specified by a  choice of modulus taking values in the coset space
$O(d-1,d-1)/O(d-1)\times O(d-1)$ identified under the T-duality
group, and can be represented locally by a choice of metric $g$
and $B$-field on the torus $T^{d-1}$, which can be combined into
$E=g+B$. There is a natural action of $O(d-1,d-1)$ on the moduli
space (acting on $E$ through fractional linear transformations).
In the twisted reduction, the moduli depend on the coordinate $y$
through the $O(d-1,d-1)$ transformation (\ref{gy}),  giving a
bundle of conformal field theories over $S^1$ with moduli $E(y)$.
This is also a bundle over the doubled torus in which there is no
dependence on the dual coordinate $\tilde y$, so that the bundle
over the dual circle is trivial. Now, in this doubled picture one
can simply choose the other polarisation, so that $\tilde y$ is
now viewed as the spacetime coordinate, and $y$ is the dual
coordinate. In this polarisation, there is no twisting on the
spacetime circle, but there is dependence on the dual coordinate
through (\ref{gy}). This is of course non-geometric, and cannot be
understood in the spacetime alone. On relabelling $y
\leftrightarrow \tilde y$, one obtains the T-dual picture in which
$y$ is the spacetime coordinate but there is a twist on the dual
circle   through (\ref{gyt}). From the point of view of conformal
field theory, it is just as natural to introduce a $\tilde
y$-twist as it is to introduce a $ y $ twist.

The general twisted construction is to introduce both a  $y$-twist
and a $\tilde y $ twist, so that the dependence of the moduli
$E(y,\tilde y)$ on $y$ and  $\tilde y $ is through an $O(d-1,d-1)$
transformation
\be { g(y, \tilde y)= \exp \left( \frac{M y}{2\pi R} \right) } {
\exp \left( \frac{ \tilde M\tilde y}{2\pi  \tilde R} \right) }
\label{gytt} \ee
where the mass-matrices $M, \tilde M$ are required to commute and
the monodromies
\be
{ {\cal M} (g)= \exp M}\qquad {  \tilde{\cal M} (g)= \exp  \tilde M }
\ee
are both in $O(d-1,d-1,\Z)$. If $g$ depends just on $\tilde y$, it
is T-dual to a T--fold construction in which the twist depends
only on $y$, so that it is locally geometric, consisting of local
spacetime patches glued together using T-dualities, gauge
transformations and diffeomorphisms. However, the general case in
which it depends on both $y$ and $\tilde y$ will not be T-dual to
anything that is even locally geometric, as there is dependence on
both $y$ and $\tilde y$.

In special cases in which the twists are elliptic, there will be
fixed points in moduli  space at which dependence on $y, \tilde y$
drops out, giving an orbifold with shifts in both $y$ and $\tilde
y$. The generalised twisting over the $\tilde y$ circle thus
provides a way of moving away from the  orbifold point in moduli
space.

There is also the possibility of intermediate cases, with special
points in moduli  space in which the dependence on $\tilde y$,
say, drops out but the dependence on $y$ does not, so that the
result can be thought of as a twisted reduction with twist
(\ref{gy}) orbifolded by the action of $\tilde {\cal M}$ together
with a shift in the $\tilde y$ direction. For example, with $d=3$,
$O(d-1,d-1)$ is locally $SL(2)\times SL(2)$, and if the $\tilde y$
twist is an elliptic element of one $SL(2)$ and the $y$ twist is
say a parabolic twist in the other $SL(2)$, there will be fixed
points in the moduli space at which the $\tilde y$ dependence
drops out in this way, but there will be no points at which the
$y$-dependence drops out.

\section{String Field Theory and Doubled Geometry}\label{sft}

   \subsection{String Field Theory}

A proper framework for thinking about this generalized T-duality
is string field theory, which was developed for toroidal compactification in
\cite{Kugo:1992md}. One can expand a string field $|\Psi\rangle$
in a basis given by the first quantized string theory defined by a
conformal field theory. If the conformal field theory has a circle
factor then a state in the first-quantized Hilbert space is
labelled as $|m, w,x, I\rangle$ where $m$ and $w$ are the momentum
and winding along the circle as above, $x^\mu$ are coordinates for
the  remaining dimensions and $I$ denotes generically all other
indices labeling the state. The string field would then have an
expansion $|\Psi\rangle^R = \sum \Psi_{m,w, I}^R(x) | m, n,x
I\rangle^R$ where the superscript denotes that the string field is
defined at radius $R$ of the circle (and the summation includes an
integration over $x$). Each component $\Psi_{m, w, I}(x)$ in this
expansion is a spacetime field. Now, clearly one expects that the
string field $|\Psi\rangle^R$ defined at a radius $R$ is T-dual to
the string field $|\Psi\rangle^{\tilde R}$ defined at the T-dual
radius ${\tilde R} = 1/R$ with Buscher-like T-duality rules that
will relate the fields $\Psi_{m,w, I}^R(x)$ to the fields
$\Psi_{{\tilde m}, {\tilde w}, I}^{\tilde R}(x)$. Note that the
Buscher rules apply to  only the lowest components of this tower
of states $\Psi_{0,0, I}$ but clearly a generalization does exist
since T-duality is an exact map that relates the Hilbert spaces of
the first quantized string theories at radius $R$ and $\tilde R$.
Given the rules of how the basis vectors $| m, n, I\rangle^R$ are
related to $| {\tilde m}, {\tilde w}, I\rangle^{\tilde R}$, which
follow from the T-duality map of the circle
conformal field theory, one knows how the fields $\Psi_{m,w, I}^R$
are related to $\Psi_{{\tilde m}, {\tilde w}, I}^{\tilde R}$.
These are the rules that must be used in doing the T-duality along
the circle $y$ in the situation at hand.

Instead of using the basis of states $|m, w, x,I\rangle$, we can
Fourier transform  to a position-basis of states $|y, \tilde y,x,
I\rangle$ depending on the periodic coordinates $y, \tilde y$.
More generally, for configurations with a $T^d$ fibration, the
position basis will depend on positions $x^\mu$ in the
non-toroidal directions, coordinates $y^i$ ($i=1,...,d$) on $T^d$
and dual coordinates $ \tilde y_i$, providing a  basis of states
$|x^\mu ,y^i, \tilde y_i, I\rangle$, and expanding a string field
will then give an infinite series of fields $\phi _I(x^\mu ,y^i,
\tilde y_i)$ (with a field corresponding to each element in a
basis of the oscillator Fock space) depending on the dual
coordinates $\tilde y$, as well as $x,y$. The $x,y,\tilde y$ are
coordinates for the   \lq doubled space' used in
\cite{Hull:2004in}, and the infinite set of fields are then fields
$\phi (x^\mu ,y^i, \tilde y_i)$ on  this doubled space. T-duality
transformations can then take a conventional field depending only
on $x,y$ to one depending on $x,\tilde y$, or to ones depending on
$x,y,\tilde y$, and the general solutions of the string field
theory equations will give fields depending on $\tilde y$, as well
as $x,y$.

\subsection{Doubled Geometry}

String field theory then leads us to   consider fields on the
doubled geometry  with coordinates $x^\mu ,y^i, \tilde y_i$, and
in \cite{Hull:2004in} it was shown how to formulate string theory
as a constrained sigma-model with target space given by this
doubled geometry. In this section, we discuss this geometry  for
the case of reductions with duality twists and their T-duals.
Reduction on $T^d$ followed by a reduction on $S^1$ with a
geometric  twist with monodromy in the group $SL(d,\Z)$ of large
diffeomorphisms of $T^d$ can be understood as a compactification
on a $T^d$ bundle over $S^1$ \cite{Hull:1998vy}. Reduction on
$T^d$ followed by a reduction on $S^1$ with a non-geometric  twist
with monodromy in the T-duality group $O(d,d;\Z)$ on $T^d$ gives a
T-fold (which of course is not a manifold in general)  but the
doubled geometry  is a bundle with fibres given by the doubled
torus $\bar {T}^{2d}$ with coordinates $y,\tilde y$. This is a
manifold, as $O(d,d;\Z)\subset SL(2d;\Z)$ acts geometrically on
the doubled torus. The base space is a circle with coordinate $u$,
say, and strings on this circle can be formulated in terms of a
doubled circle $\bar T^2$ with coordinates $u,\tilde u$. A
generalised T-duality on the $u$-circle takes a $\bar {T}^{2d}$
bundle  over the $u$-circle to a $\bar {T}^{2d}$ bundle  over the
dual  $\tilde u$-circle, corresponding to a reduction with twist
over the $\tilde u$-circle, and the general non-geometric
reduction of this type will correspond to a doubled geometry which
is a bundle with doubled fibres  $\bar {T}^{2d}$   with
coordinates $y,\tilde y$ and a doubled base $\bar T^2$ with
coordinates $u,\tilde u$, corresponding to a   reduction with
twists on both the $u$-circle and the dual  $\tilde u$-circle.
This doubled construction is closely related to what has become
known as  \lq generalised geometry' (see e.g.
\cite{Hitchin:2004ut}), and in fact can be thought of as a further
generalisation of this \cite{Hull:2004in}.  This geometry will be
explored further elsewhere.

\section{Discussion}\label{discussion}

We have seen that general non-geometric backgrounds have
dependence both on the  conventional coordinates $y$ and dual
coordinates $\tilde y$, and that fields in general depend on both
$y$ and  $\tilde y$. The extension of T-duality to circles for
which the natural $U(1)$ action is not  isometric necessarily
leads to such backgrounds. In general, one would expect momentum
modes to be sensitive to the $y$-dependence of a background, while
momentum modes would be sensitive to the $\tilde y$-dependence, so
that different probes would \lq see' different backgrounds. A world-sheet approach to
extending  T-duality to backgrounds without isometries was proposed in \cite{Evans:1995su}, and it would be interesting to compare this with our construction.

An important class of backgrounds arise from reductions with
duality twists,  with monodromy {\cal M} around a circle with
coordinate $y$, and we have considered the example in which this
is a T-fold corresponding to a bundle whose fibres are CFT's on
$T^n$, with moduli depending on $y$. A T-duality $T$ on the $T^n$
fibres takes this to a twisted reduction with conjugate monodromy
$T{\cal M}T^{-1}$. We propose that  a generalised  T-duality on
the base circle of the original background with coordinate $y$
takes this to a twisted reduction over the dual circle, with
moduli now depending on the dual coordinate $\tilde y$, with
monodromy {\cal M} on the  $\tilde y$ circle. In the example of a
$T^3$ with constant flux, one T-duality takes this to a $T^2$
bundle over $S^1$ with parabolic twist, a further T-duality in a
fibre direction takes this to a T-fold with monodromy   ${\cal
M}\in O(2,2;\Z)$ and finally a third T-duality on the base circle
takes this to a bundle of $T^2$ CFT's over the dual circle with
the same  monodromy   ${\cal M}\in O(2,2;\Z)$.

An interesting example of   geometries with dependence on a dual
coordinate has already appeared in the literature
\cite{Gregory:1997te, Tong:2002rq, Harvey:2005ab, Okuyama:2005gx}. A Kaluza-Klein
monopole solution of string theory (given by the product of
Euclidean Taub-NUT with 6-dimensional Minkowski space) is T-dual
to a NS 5-brane, smeared over one of the four transverse
directions,  which is compactified to a circle \cite{Hull:1994ys}.
A localised version of the NS 5-brane compactified on a circle is
obtained by taking an infinite periodic array of parallel NS
5-branes along a line, and then periodically identifying that
line. This now has non-trivial dependence on the coordinate $y$ of
that circle, but is independent of the dual coordinate $\tilde y$.
On T-dualising, one would expect $\tilde y$ to become the
coordinate $x$ on the $S^1$ fibre of the Kaluza-Klein monopole,
and $y$ to become the dual coordinate $\tilde x$. In
\cite{Gregory:1997te, Tong:2002rq, Harvey:2005ab}, evidence was
given that the T-dual of the localised NS 5-brane should be a
version of the Kaluza-Klein monopole with non-trivial dependence
on the dual coordinate $\tilde x$, arising from the dependence of
the NS 5-brane on $y$, so that whereas momentum modes   experience
a throat-like geometry in the 5-brane, string winding modes feel a
stringy throat-like structure of the Kaluza-Klein monopole.
Non-trivial $\tilde y$ dependence  of the Kaluza-Klein monopole
can be understood as arising from world-sheet instantons
\cite{Gregory:1997te}, and this  has been verified by linear
sigma-model calculations \cite{ Tong:2002rq, Harvey:2005ab}. It
was also argued in   \cite{Gregory:1997te} that this T-duality
might be understood in terms of a T-duality of higher modes of the
string field.

Conventional effective supergravity theory in 10-dimensions is clearly
inadequate for  the study of general non-geometric string
backgrounds, although the effective supergravity theory arising
from reduction on a  non-geometric   background does provide
useful information. There is clearly much that needs to be done to
obtain a better understanding of the doubled geometry arising in
this context, and the generalisations involving U-duality and
brane wrapping modes.

\section*{Acknowledgements}

We would like to thank  ICTP and the   Institute for Mathematical
Sciences, Imperial College for hospitality and support. This work
was partially supported by an EPSRC visiting fellowship.

\end{document}